\documentclass[12pt,preprint]{aastex}

\bibpunct{(}{)}{;}{a}{}{,}

\newcommand{\kms}{km\,s$^{-1}$}
\newcommand{\fuse}{\emph{FUSE}}
\newcommand{\visir}{\emph{VISIR}}

\newcommand{\iue}{\emph{IUE}}

\newcommand{\flux}{ergs\,s$^{-1}$\,cm$^{-2}$}

\newcommand{\teff}{T$_{\rm eff}$}

\newcommand{\Av}{A$_{\rm v}$}

\newcommand{\myemail}{claire.martin-zaidi@cea.fr}
\shorttitle{Detection of H$_2$ in the HD97048 disk}
\shortauthors{Martin-Za{\"\i}di et al.}

\begin{document}


\title{Detection of warm molecular hydrogen in the circumstellar disk
  \\ around the Herbig A\lowercase{e} star HD97048}

\author{C. Martin-Za{\"\i}di\altaffilmark{1},
  P-.O. Lagage\altaffilmark{1}, E. Pantin\altaffilmark{1} and
  E. Habart\altaffilmark{2}}

\altaffiltext{1}{Laboratoire AIM, CEA/DSM - CNRS - Universit\'e Paris
  Diderot, DAPNIA/Service d'Astrophysique, Bat. 709, CEA/Saclay, 91191
  Gif-sur-Yvette Cedex, France, \myemail,
  pierre-olivier.lagage@cea.fr, epantin@cea.fr}

\altaffiltext{2}{Institut d'Astrophysique Spatiale, 91405 Orsay,
  France, emilie.habart@ias.u-psud.fr}

\begin{abstract}
  We present high resolution spectroscopic mid-infrared observations
  of the circumstellar disk around the Herbig Ae star HD97048 with the
  {\it VLT Imager and Spectrometer for the mid-InfraRed} ({\it
    VISIR}). We detect the S(1) pure rotational line of molecular
  hydrogen (H$_2$) at 17.035 $\mu$m arising from the disk around the
  star. This detection reinforces the claim that HD97048 is a young
  object surrounded by a flared disk at an early stage of evolution.
  The emitting warm gas is located within the inner 35 AU of the disk.
  The line-to-continuum flux ratio is much higher than expected from
  models of disks at local thermodynamics equilibrium. We investigate
  the possible physical conditions, such as a gas-to-dust mass ratio
  higher than 100 and different excitation mechanisms of molecular
  hydrogen (X-ray heating, shocks, ...)  in order to explain the
  detection. We tentatively estimate the mass of warm gas to be in the
  range from 10$^{-2}$ to nearly 1 M$_{Jup}$.  Further observations
  are needed to better constrain the excitation
  mechanisms as well as the mass of gas.  \\
\end{abstract}


\keywords{stars: pre-main sequence -- stars: individual (HD97048) --
  (stars:) circumstellar matter -- (stars:) planetary systems:
  protoplanetary disks -- infrared: stars}



\section{Introduction}

Circumstellar (CS) disks surrounding pre-main sequence stars are
supposed to be the location of planet building. The characterization
of the gaseous component, which initially represents 99\%\ of the
total disk mass, is a key research question towards an understanding
of protoplanetary disks and planet formation. However, from previous
observations, little is known about the gas compared to the
dust. Major questions concerning planet formation remain. How massive
are the disks? Can giant planets form in every disk?  How long does
the planet formation process take?  Detailed information is required
about the gas in disks in order to address these questions. In
particular, characterizing the warm gas phase in the inner disk
(R$<$50 AU), where planet formation is supposed to take place, is an
essential step.

Molecular hydrogen (H$_2$) is the main constituent of the molecular
cloud from which the young star is formed and is also expected to be
the main component of the CS disk.  It is expected to be at least
10$^4$ times more abundant than other gas tracers such as carbon
monoxide (CO) \citep[e.g.][]{Bell06}, since it self-shields very
efficiently against photodissociation by far-ultraviolet (FUV) photons
and does not freeze effectively onto grain surfaces. H$_2$ is the only
molecule that can directly constrain the mass reservoir of warm and
hot molecular gas in disks. Indeed, the detection of H$_2$ excited by
collisions allows us to measure the temperature and density of the
warm gas. Unfortunately, direct observation of H$_2$ is
difficult. Electronic transitions occur in the ultraviolet to which
the Earth's atmosphere is opaque, and rotational and ro-vibrational
transitions at infrared (IR) wavelengths are faint because of their
quadrupolar origin. FUV absorption lines have been observed with the
\fuse\ satellite in the spectrum of some Herbig Ae stars (HAes) and
show the presence of warm (T$_{gas} > 300$ K) molecular hydrogen gas
in the CS environment of these stars \citep{klr07b}. However,
absorption observations require specific configurations to observe the
gas within the disks, i.e. nearly edge-on. Due to the high inclination
angles to the lines of sight estimated for the disks observed by
\citet{klr07b}, the detected H$_2$ is not in the disk. Those authors
concluded that the lines of sight probably pass through a thin layer
of warm/hot gas above the surface of the disk which is produced by the
photoevaporation of the disk. Searches for mid-IR H$_2$ rotational
emission lines have been performed using different space- and
ground-based instruments. Space-based instruments on ISO and Spitzer
have low spectral and spatial resolution, and therefore have not
yielded an unambiguous detection of H$_2$ line emission from
protoplanetary disks \citep{Thi01, Pascucci06}. Indeed, when
detections of H$_2$ towards a few pre-main sequence stars have been
claimed with ISO-SWS \citep{Thi01}, ground-based observations showed
that contamination from surrounding cloud material was important and
that ISO detections were likely not dominated by the emission of the
disk gas \citep[e.g.][]{Richter02, Sako05}. Recently,
\citet{Carmona07} estimated the line-to-continuum ratio that should be
observed for H$_2$ transitions in the mid-IR. They used a two-layer
model \citep{Chiang97, DULLEMOND01} of a gas-rich disk (column density
of N(H$_2$)=10$^{23}$ cm$^{-2}$) seen face-on, located at 140 pc from
the Sun, with local thermodynamics equilibrium (LTE) for the gas and
dust, T$_{gas}$=T$_{dust}$, and assuming a gas-to-dust mass ratio
about 100. Those authors concluded that the expected peak flux of the
S(1) line at 17.035 $\mu$m, observed at a spectral resolution of
20\,000, should be less than 0.3\%\ of that of the continuum at
temperatures higher than 150 K, and thus should not be observable with
the existing instruments. Indeed, they did not detect any H$_2$ mid-IR
emission line in their sample of 6 Herbig Ae stars.

However, H$_2$ rotational lines have been recently detected in the
disk around one Herbig Ae star, namely AB Aur, with the high spectral
and spatial resolution TEXES spectrometer \citep{Bitner07}. These
detections imply that H$_2$ can be observed in the mid-IR domain when
particular physical conditions exist in disks.

The {\it VLT Imager and Spectrometer for the mid-InfraRed}
\citep[\visir;][]{Lagage04} has the high spectral ($10\,000 < R <
30\,000$) and spatial resolution necessary to pick up such narrow gas
lines from the disks. In addition, high spectral resolution is a key
element to disentangle the H$_2$ line from the absorption lines due to
the Earth's atmosphere. The spectral ranges covered by \visir\ offer
access to the most intense pure rotational lines of molecular hydrogen
(H$_2$): S(1) ($v=0-0, J=3-1$) at 17.0348 $\mu$m, S(2) ($v=0-0,
J=4-2$) at 12.2786 $\mu$m, S(3) ($v=0-0, J=5-3$) at 9.6649 $\mu$m, and
S(4) ($v=0-0, J=6-4$) at 8.0250 $\mu$m. The S(0) ($v=0-0, J=2-0$)
transition near 28 $\mu$m is not observable from the ground due to the
Earth's atmospheric absorption.

A particularly interesting object to study the CS material around a
pre-main sequence intermediate mass star is HD97048. HD97048 is a
nearby, relatively isolated Herbig A0/B9 star located in the Chameleon
cloud at a distance of 180 pc \citep{VdA98}. Its age has been
estimated from evolutionary tracks to be of the order of 3 million
years (kindly computed by L. Testi and A. Palacios). This star is
known to be surrounded by an extended CS disk. The \visir\ imaging
observations of this star conducted in 2005 June 17 and 19, have
revealed an extended emission of PAHs (Polycyclic Aromatic
Hydrocarbons) at the surface of a flared disk inclined to the line of
sight by 42.8$_{-2.5}^{+0.8}$ degrees \citep{Lagage06}.  This is the
only Herbig star for which the flaring of the disk has been observed
by direct imaging. The flaring index has been measured to be
1.26$\pm$0.05, in good agreement with hydrostatic flared disk models
\citep{Lagage06, Doucet07}. This geometry implies that a large amount
of gas should be present to support the flaring structure and that the
disk is at an early stage of evolution. This star is thus one of the
best candidates to study the gas component in the disks of HAe stars.

In this paper, we present \visir\ observations of the S(1) pure
rotational emission line of molecular hydrogen at 17.03 $\mu$m arising
from the disk of HD97048.

\section{Observations and data reduction}

HD97048 was observed for 1800s with the high spectral resolution
long-slit mode of \visir\ in June 22, 2006. The central wavelength of
the observation was set to 17.035 $\mu$m. We used the 0.75'' slit,
providing a spectral resolution about 10\,000, i.e.  $\Delta v$=30
\kms.

\begin{figure}[!htbp]
\centering
\includegraphics[width=8cm]{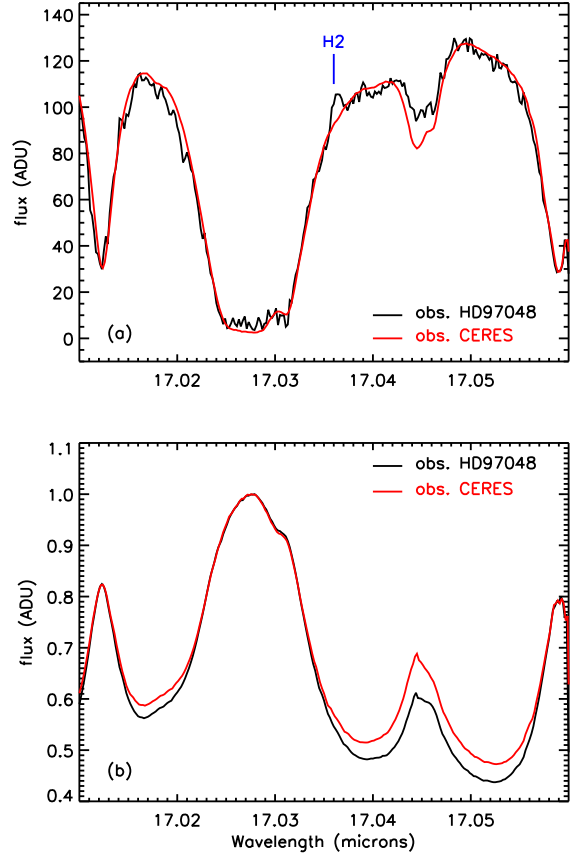}
\caption{{\it (a):} \visir\ spectrum of the CERES asteroid (red)
  overplotted on the HD97048 spectrum (black). {\it (b):} Spectrum of
  the sky during the observations of HD97048 and CERES.}
\label{CERES}
\end{figure}

The weather conditions were very good and stable during the
observations; the optical seeing was less than 0.66'' and the airmass
($<$1.8) was close to the minimum airmass accessible when observing
this object from the Paranal ESO observatory. The standard "chopping
and nodding" technique was used to suppress the large sky and
telescope background dominating at mid-infrared wavelengths.
Secondary mirror chopping was performed in the North-South direction
with an amplitude of 8'' at a frequency of 0.025 Hz. Nodding
technique, necessary to compensate for chopping residuals, was applied
using a telescope offset of 8'' in the South direction, every 3
minutes. The pixel scale is 0.127''/pixel resulting in a total field
of view along the slit about 32.5''. The elementary frames were
combined to obtain chopping/nodding corrected data. \visir\ detector
is affected by stripes randomly triggered by some abnormal high-gain
pixels. A dedicated destriping method has been developped to suppress
them (Pantin 2007, in preparation). In order to correct the spectrum
from the Earth's atmospheric absorption and obtain the absolute flux
calibration, we observed the CERES asteroid and the standard star
HD89388 (see www.eso.org/VISIR/catalogue) just before and after
observing HD97048. HD89388 and CERES were observed at nearly the same
airmass and seeing conditions as the object. As shown in
Fig.~\ref{CERES}{\it b}, airmasses are slightly different between the
observation of CERES and that of HD97048. However, the discrepancy
between the two spectra of the sky cannot be responsible for the
emission feature we observe in the HD97048 spectrum around
17.035$\mu$m, i.e. the H$_2$ line (Fig.~\ref{CERES}{\it a}). We thus
have divided the spectrum of HD97048 by that of CERES (which has a
much better signal-to-noise ratio than that of the standard star
HD89388) to correct for the telluric absorption, and used the HD89388
observed and modelled spectra \citep{Cohen99} to obtain the absolute
flux calibration. The wavelength calibration is done by fitting the
observed sky background features with a model of the Paranal's
atmospheric emission.

We note that \citet{Valenti00} found \Av=0.24 mag from the fit of the
\iue\ spectrum of HD97048, thus we have not corrected the spectrum for
dust extinction, since it is neglectible in our wavelength range for
\Av$< 40$ mag \citep{Fluks94}.

\section{Results}

\begin{figure}[!h]
\centering
\includegraphics[width=8cm]{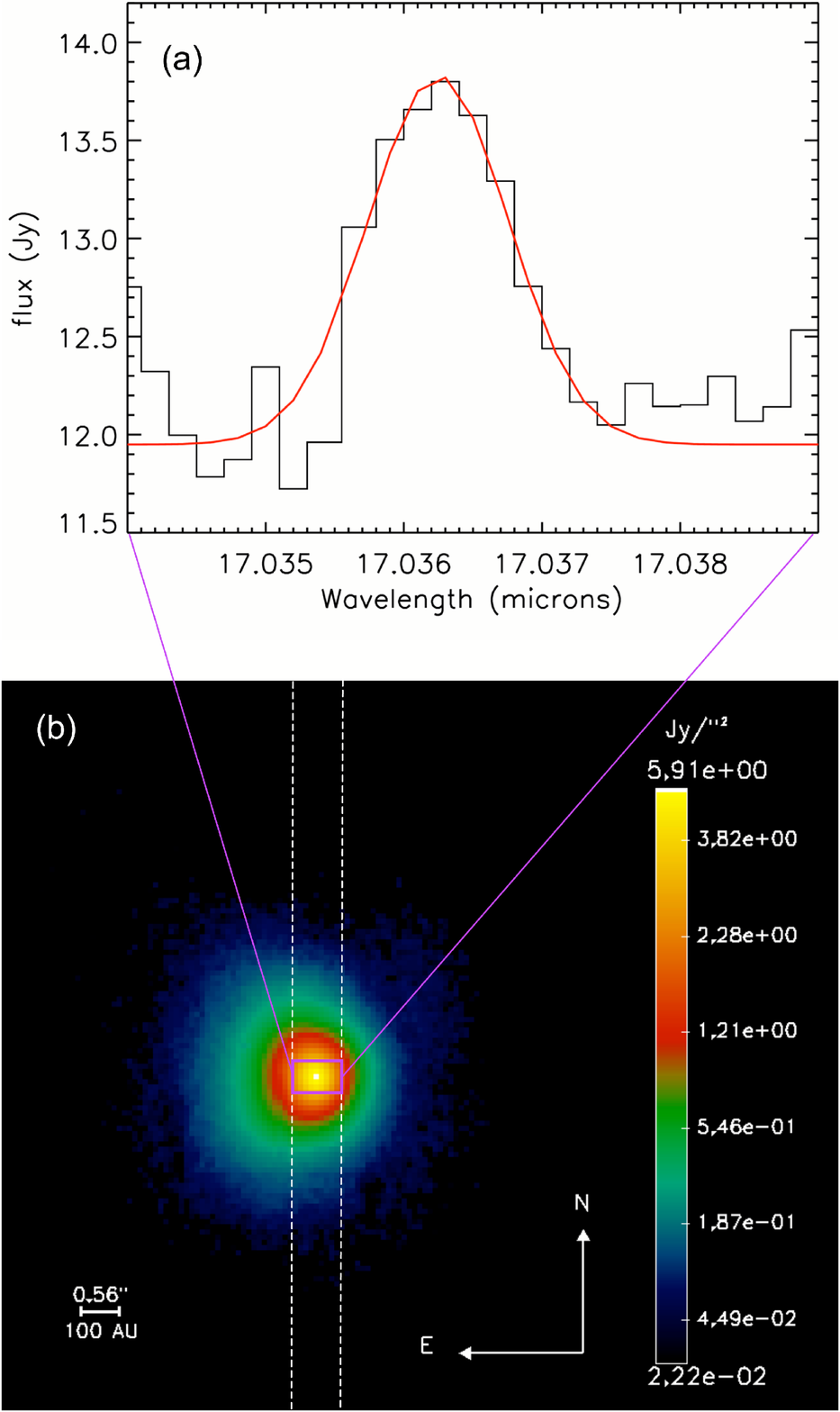}
\caption{{\it (a)}: S(1) H$_2$ emission line from HD97048 disk,
  observed with the high spectral resolution mode of \visir. Black
  line: observed spectrum corrected for telluric absorption. Red line:
  fit of a gaussian with FWHM equal to a spectral resolution
  element. Here the spectrum has not been corrected for the Earth's
  rotation. {\it (b):} \visir\ false-color image of the emission from
  the CS material surrounding HD97048 \citep{Lagage06}. White dotted
  lines: position of the 0.75'' slit of the spectrograph during the
  observation of H$_2$ emission. Purple rectangle: one element of
  spatial resolution (diffraction limited). As the line is not
  spatially resolved, the H$_2$ emission arises within this element.}
\label{raie}
\end{figure}

As shown in Fig.~\ref{raie}{\it a}, we have detected the H$_2$ pure
rotational S(1) line near 17.03 $\mu$m. In the flux-calibrated
spectrum, the standard deviation ($\sigma$) of the continuum flux was
calculated in regions less influenced by telluric absorption, and
close to the feature of interest. We deduced a 6$\sigma$ detection in
amplitude for the line, corresponding to a signal-to-noise of about
11$\sigma$ for the line, when integrating the signal over a resolution
element (6 pixels). The line is not resolved as we can fit it with a
Gaussian with a full width at half maximum equal to a spectral
resolution element of 30 \kms\ (see Fig.~\ref{raie}{\it a}).  From our
fit, assuming the emission arises from an isothermal mass of optically
thin H$_2$, we derived an integrated flux in the line of
2.4$\times$10$^{-17}$ W m$^{-2}$ or 2.4$\times$10$^{-14}$ \flux.

Once the spectrum is corrected from the Earth's rotation, and knowing
the heliocentric radial velocity of HD97048 \citep[+21
\kms;][]{Acke05}, we estimated, from the wavelength position of the
Gaussian peak, the radial velocity of H$_2$ to be about 4$\pm$2 \kms\
in the star's rest frame.  We thus considered that the radial velocity
of the H$_2$ is similar to that of the star, implying that the
emitting gas is bound to the star. The H$_2$ line is not resolved
spatially. Given the \visir\ spatial resolution of about 0.427'' at
17.03 $\mu$m, and the star distance (180 pc from the sun), we can
assess that the emitting H$_2$ is located within the inner 35 AU of
the disk (Fig.~\ref{raie}{\it b}). We estimated the corresponding
column densities and masses as a function of prescribed temperatures
(Table~\ref{tab1}). For this purpose, we first assumed that the line
is optically thin and that the radiation is isotropic. In this
context, assuming that the first rotational levels (up to $J=3$) of
H$_2$ are thermalized, and thus that their populations follow the
Boltzmann law for a given temperature (LTE), the column densities are
derived from the following formula:

\begin{eqnarray}
I_{ul}=\frac{hc}{4 \pi \lambda} N_u(H_2) A_{ul} ~~~~
ergs\,s^{-1}\,cm^{-2}\,sr^{-1}
\end{eqnarray}

{\noindent}where $I_{ul}$ is the integrated intensity of the line,
$\lambda$ is the wavelength of the transition $J=u-l$, $A_{ul}$ is the
spontaneous transition probability, $N_u(H_2)$ is the column density
of the upper rotational level of the transition. Since the line is not
spatially resolved, we calculated a lower limit on $I_{ul}$ by
dividing the integrated flux value by the solid angle of the point
spread function, and thus obtained lower limits on the total column
densities. Under the same assumptions as used for the calculations of
the column densities and assuming that the medium we observe is
homogeneous, the mass of warm H$_2$ is given by \citep{Thi01}:

\begin{eqnarray}
M_{gas}=f\times 1.76\times 10^{-20}\frac{F_{ul} d^2}{(hc / 4 \pi
  \lambda) A_{ul} x_u(T)}~~M_{\odot}
\end{eqnarray}

{\noindent}where $F_{ul}$ is the line flux, $d$ the distance in pc to
the star, $x_u(T)$ is the fractional population of the level $u$ at
the temperature $T$ in LTE \citep[for details on the calculation
method, see][]{Van_Dishoeck_92}, $f$ is the conversion factor required
for deriving the total gas mass from the H$_2$-ortho or H$_2$-para
mass. Since M$_{H2}$=M$_{H2}$(ortho)+M$_{H2}$(para), $f$
=1+1/(ortho/para) for the S(1) line (a H$_2$-ortho transition). The
equilibrium ortho-para ratio at the temperature T was computed using
\citet{Takahashi01}.
 
We also estimated the dust mass producing the flux level of the
continuum in the spectrum. We used a simple model of optically thin
emission of a given mass of dust at the surface of a disk. The grains
have sizes between 0.01 and 100 $\mu$m and a size distribution
following a power-law with an index of -3.5. A fixed composition,
mixture of amorphous silicates (50\%) and amorphous carbon (50\%) is
assumed. For different temperatures (150, 300, 1000 K) assigned to the
dust, we computed the corresponding mass of dust and derived
gas-to-dust mass ratios. Our results are tabulated in
Table~\ref{tab1}.

\section{Discussion}

Our high resolution spectroscopic observation of the S(1) pure
rotational line of H$_2$ at 17.03 $\mu$m of HD97048 has revealed the
presence of significant amounts of warm gas in the inner 35 AU of the
disk. From a gaussian fit of the emission line, we derived very high
column densities of warm gas, which are more than two orders of
magnitude higher than those generally observed in the CS environment
of Herbig Ae stars \citep{klr07b}. This confirms that HD97048 is a
young object surrounded by a circumstellar disk at an early stage of
evolution. Indeed photoevaporation of the gas is expected to clear up
the inner part of the disk within 3 million years \citep{Takeuchi05}.

We derived masses of the warm gas in the range from 10$^{-2}$ to
nearly 1 M$_{Jup}$ (1 M$_{Jup}$ $\sim$ 10$^{-3}$ M$_{\odot}$)
depending on the adopted temperature, and assuming LTE. The masses
derived here are lower than those of \citet{Lagage06} who have
estimated a minimum mass of gas in the inner disk to be of the order
of 3 M$_{Jup}$. But it should be pointed out that mid-IR H$_2$ lines
are only probing warm gas located in the surface layer of the disk,
when a higher mass of colder gas is expected to be present in the
interior layers of the disk. In any case, the finding of warm
molecular hydrogen reinforces the claim that a large amount of cold
gas is present in the disk to support its flaring geometry
\citep{Lagage06}.

It is generally accepted that the first rotational levels ($J$) of
H$_2$ are populated by thermal collisions, an excitation mechanism
which requires kinetic temperatures higher than 150 K to produce the
S(1) transition. Assuming equal dust and gas temperatures, we
estimated dust masses responsible for the continuum emission and
derived gas-to-dust mass ratios in the range from 3260 to 14164
(Table~\ref{tab1}), much larger than the canonical value of 100.
These crude estimates are in agreement with more sophisticated models
such as two-layer LTE disk models \citep{Carmona07}. Indeed, by
scaling the gas-to-dust mass ratio found here to the canonical value
of 100, we obtained a peak line flux of about 0.46\%\ of that of the
continuum at 150 K, decreasing to 0.1\%\ of the continuum at 1000 K,
which is close to the line-to-conitnuum ratios calculated in disks
models by \citet{Carmona07}. Thus one possible interpretation of our
observation is that the dust is partially depleted from the disk
surface layer, where the H$_2$ emission originates. The spatial
decoupling between the gas and the dust may be due to dust settling or
dust coagulation into larger particles.

However, other excitation mechanism cannot be excluded. Several
competing mechanisms could contribute to the excitation of molecular
hydrogen such as UV pumping, shocks, X-rays, etc..., \citep[see review
papers by][]{Habart04b, Snow06} and could be responsible for the
observed emission. Weak X-ray emission has been detected from HD97048
by ROSAT \citep{Zinnecker94}. X-rays and UV photons are likely
candidates to heat the gas to temperatures significantly higher than
those of the dust \citep{Glassgold07} and could partly explain a high
line-to-continuum ratio. According to radiative transfer models of
disks around T Tauri stars \citep{Nomura05, Nomura07}, X-ray heating
could significantly increase the line-to-continuum flux ratio, but,
applying the same increase factor to HAes, the S(1) H$_2$ line would
still be below the detection limit of \visir.

Note that the present \visir\ observation does not allow us to
discriminate between the different possible physical origins of the
emission of H$_2$. New observations of HD97048 will be performed with
\visir\ in order to observe the other pure rotational lines of
H$_2$. The detection of these lines would help to better constrain the
temperature (and thus the mass) of the warm gas.

Our results are very similar to those obtained by \citet{Bitner07} for
AB Aur with the TEXES instrument. Indeed, those authors have shown
that the emitting warm gas is located in the inner 18 AU of the disk
around AB Aur. For the two stars, the gas has not completely
dissipated in the inner region of the disk in a lifetime of about 3
Myrs. HD97048 and AB Aur have nearly identical astrophysical
parameters (\teff, age, mass, distance). Their disks are flared
\citep{Pantin05, Lagage06} and seem to be in similar evolutionary
states, which could well be a disk old enough such that the dust
sedimenation/coagulation has already been at work, but young enough
such that the gas has not yet been photoevaporated. It is not possible
to draw definite conclusions with only two examples and it would be
interesting to observe other HAe stars similar to AB Aur and
HD97048. The high angular resolution and high spectral resolution
available with ground-based instruments are key advantages over
space-based instruments such as ISO-SWS in order to obtain firm
detections of H$_2$ from disks. \\

This work is based on observations obtained at ESO/VLT (Paranal) with
\visir, program number 077.C-0309(B). We would like to thank S. Madden
for her careful reading of the manuscript. CMZ warmly thanks C. Gry,
C. Doucet, P. Didelon and A. Carmona for fruitfull discussions.\\


\clearpage


\begin{table}[!htbp] 
\begin{center}
  \caption{Total column density and mass of H$_2$, mass of dust and
    gas-to-dust mass ratio as a function of the temperature.}
\begin{tabular}{lcccccc}
  \hline
  \hline
  &    &     150 K        &         300 K         &    1000 K \\
  \hline
  N(H$_2$) (cm$^{-2}$)  &  & $>$2.19$\times$10$^{23}$ & $>$1.41$\times$10$^{22}$ & $>$3.27$\times$10$^{21}$ \\
  M(H$_2$) (M$_{Jup}$)  &  & 7.37$\times$10$^{-1}$    & 4.51$\times$10$^{-2}$    & 1.33$\times$10$^{-2}$\\
  M$_{dust}$ (M$_{Jup}$) &  & 2.23$\times$10$^{-4}$    & 1.03$\times$10$^{-5}$    & 9.39$\times$10$^{-7}$ \\
  gas-to-dust mass ratio    &  &  3260                    &   4378                  & 14164 \\     
  \hline
\end{tabular}
\label{tab1} 
\end{center}
\end{table}

\clearpage



\clearpage

\end{document}